\documentclass[twocolumn,showpacs,preprintnumbers,amsmath,amssymb]{revtex4}
\usepackage{tabularx,graphicx}

\usepackage{color}
\usepackage{hyperref}
\hypersetup{
    colorlinks=true,
    linkcolor=blue,
    filecolor=blue,      
    urlcolor=blue,
}

\begin{document}

\newcommand{\beq}{\begin{equation}}
\newcommand{\eeq}{\end{equation}}
\newcommand{\beqn}{\begin{eqnarray}}
\newcommand{\eeqn}{\end{eqnarray}}
\newcommand{\bmath}{\begin{subequations}}
\newcommand{\emath}{\end{subequations}}
\newcommand{\bra}[1]{\langle #1|}
\newcommand{\ket}[1]{|#1\rangle}

\title{Alfven-like waves along   normal-superconductor phase boundaries}
\author{J. E. Hirsch }
\address{Department of Physics, University of California, San Diego,
La Jolla, CA 92093-0319}

\begin{abstract} 
Alfven waves are  transverse magneto-hydrodynamic waves resulting from motion of a conducting fluid in direction perpendicular to an applied
magnetic field, that propagate along the magnetic field direction. I propose that  Alfven-like waves can propagate along   normal-superconductor 
phase boundaries in the presence of a magnetic field.  
This requires charge flow and backflow across the normal-superconductor phase boundary when the boundary moves, which is predicted by the theory of hole 
superconductivity but not by   the conventional theory of superconductivity. The  magnetic field and the  domain wall energy provide elasticity, and the normal charge carriers' effective mass  provides inertia. It is essential that the normal state charge carriers are holes. I propose an 
experimental search for Alfven-like waves in superconductors.  \end{abstract}
\maketitle

\section{introduction}
In a perfectly conducting fluid, magnetic field lines are frozen into the fluid and move together with the fluid
 (Alfven's theorem) \cite{davidson}. 
If the fluid is initially at rest in a uniform magnetic field and a small velocity transverse to the magnetic field  is induced, 
it acts as a source for propagation of a transverse magneto-hydrodynamic wave in the direction of the magnetic field, 
called an Alfven wave \cite{alfven}. 
One may think of the magnetic field lines as  strings in tension, the transverse fluid  motion drags the magnetic field lines
and creates curvature in them, and  they
 will snap back as if they were elastic strings, dragging the  fluid along. The backward motion overshoots due to the fluid's inertia and oscillatory motion results, just like with  a massive elastic string.
Figure 1 shows qualitatively the physics that is involved.

        \begin{figure} [h]
 \resizebox{6.5cm}{!}{\includegraphics[width=6cm]{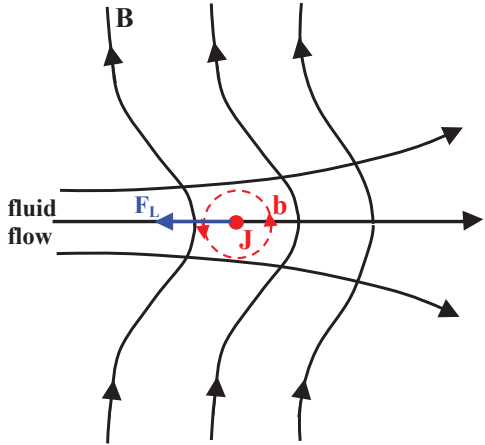}} 
 \caption {Qualitative description of Alfven waves. In a conducting fluid flowing to the right in the presence of a vertical
 magnetic field $B$, a current $J$ flowing
 out of the paper is induced due to the magnetic Lorentz force. The magnetic field $b$ created by this
 current (dashed red circle) modifies the field lines giving them curvature. The Ampere
 force $F_L$ acts on the current $J$ exerting a force to the left, causing the fluid to
  flow back after reaching a maximum amplitude.   }
 \label{figure1}
 \end{figure} 

Consider a type I superconductor in the presence of a magnetic field. One can have phase equilibrium between normal 
and superconducting phases \cite{londonh}, where magnetic field lines and supercurrents
  flow parallel to the phase boundary and perpendicular to each other, penetrating a London penetration depth into the superconducting region. Whenever possible the field lines and the phase boundary will be straight, to minimize  domain wall
energy and magnetic energy. We may ask: if a perturbation is set up that gives curvature to the domain wall and the magnetic field lines, will this be a source
of Alfven-like waves propagating parallel to the phase boundary?

Within the conventional theory of superconductivity there is no  dynamical description of how the phase boundary between
normal and superconducting phases moves in the transition between normal and superconducting states in a magnetic field
(Meissner effect) or its reverse, the superconductor-normal transition. In particular, there is no dynamical explanation for how  the supercurrent starts and stops and
how momentum is conserved \cite{ondyn}. We have recently proposed a description of these processes
\cite{disapp,revers,momentum,entropy,whyholes} using physical elements
derived from the theory of hole superconductivity \cite{holesc} that are not part of the conventional theory. Here we argue that this physics  allows for propagation of
Alfven-like waves along normal-superconductor phase boundaries, that should be experimentally detectable. We also argue
that no such   wave propagation should occur within the conventional theory of superconductivity.

        \begin{figure} [h]
 \resizebox{8.5cm}{!}{\includegraphics[width=6cm]{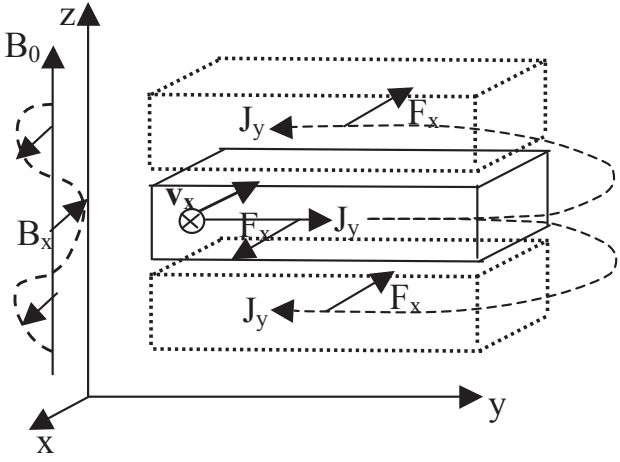}} 
 \caption {The magnetic field $B_0$ points in the $+z$ direction.
  In the center rectangular prism, fluid moves in the $-x$ direction and current is generated in the $+y$ direction. 
 The resulting patterns of currents, forces and magnetic field are explained in the text.   }
 \label{figure1}
 \end{figure} 
 
\section{alfven waves}
We start by  reviewing the derivation of Alfven waves in a simple geometry \cite{alfven2}. Consider  
a perfectly conducting incompressible charge-neutral fluid with a uniform
magnetic field $B_0$ pointing in the z direction, as shown in Fig. 2. Assume a slab of fluid (solid prism) moves in the negative $x$ direction
with speed $v_x$. In the frame of the moving fluid an electric field in the $y$ direction arises, of magnitude $(v_x/c)B_0$, that
generates a current $J_y$ in the $y$  direction. The Lorentz force acting on this current $F_x$ points in the positive $x$ direction,
thus opposing the motion of the fluid. The current in this slab returns flowing in the opposite direction above and
below the slab (dotted slabs), and on these currents the force $F_x$ points in the negative $x$ direction as shown in the figure. 
These current flows modify the initial magnetic field $B_0\hat{z}$ by adding small magnetic field components $B_x$ in the
$\pm x$ direction as shown to the left of the $z-$ axis in the figure. Thus, the fluid and the magnetic field lines undergo
oscillatory motion in the x-direction  and an oscillatory electric current flows in the $y$ direction.

For simplicity we assume translational invariance in the $y$ and $x$ directions, hence we have
$\vec{B}=(B_x(z,t),0,B_0)$, $\vec{J}=(0,J_y(z,t),0)$, $\vec{v}=(v_x(z,t),0,0)$, $\vec{E}=(0,E_y(z,t),0)$. 
$\vec{E}$ is the electric field in the lab frame that appears due to Faraday's law.
 From Ampere's law,
\beq
\vec{\nabla}\times\vec{B}=\frac{4\pi}{c}\vec{J}==>\frac{\partial B_x}{\partial_z}=\frac{4\pi}{c}J_y
\eeq
The current is given by
\beq
\vec{J}=\sigma(\vec{E}+\frac{\vec{v}}{c}\times\vec{B})
\eeq
with $\vec{E}$ the electric field, so for infinite conductivity $\sigma$
\beq
\vec{E}=-\frac{\vec{v}}{c}\times\vec{B}==>E_y=\frac{v_x}{c}B_0 .
\eeq
Faraday's law yields
\beq
\vec{\nabla}\times\vec{E}=-\frac{1}{c}\frac{\partial \vec{B}}{\partial t}==>\frac{\partial E_y}{\partial z}=\frac{1}{c}\frac{\partial B_x}{\partial t}
\eeq
and the equation of motion for the fluid, assuming zero viscosity, is
\beq
\rho \frac{\partial \vec{v}}{\partial t}=-\vec{\nabla} p+\frac{\vec{J}}{c}\times \vec{B}==>\rho \frac{\partial v_x}{\partial t}=\frac{J_y}{c}B_0
\eeq
to linear order, where $p(z,t)$ is the pressure and $\rho$ the mass density. From Eqs. (1) and (5)
\beq
\frac{\partial B_x}{\partial z}=\frac{4\pi \rho}{B_0}\frac{\partial v_x}{\partial t} .
\eeq
On the other hand from Eqs. (3) and (4)
\beq
v_x=\frac{c}{B_0}E_y==>\frac{\partial v_x}{\partial z}=\frac{1}{B_0}\frac{\partial B_x}{\partial t}
\eeq
and combining Eqs. (6) and (7)   we obtain
\beq
\frac{\partial^2 B_x}{\partial z^2}=\frac{4\pi \rho}{B_0^2}\frac{\partial ^2 B_x}{\partial t^2}
\eeq
and a similar equation for $v_x$, describing dispersionless waves that propagate in the $z$ direction with speed
\beq
v_A=   \sqrt{\frac{B_0^2}{ 4\pi \rho} }
\eeq
the Alfven speed. The magnetic energy density $B_0^2/8\pi$ gives the elasticity, and the fluid mass density $\rho$ provides the inertia,
both required for wave propagation with speed $v\sim\sqrt{elasticity/inertia}$ as in a stretched string or a compressible medium. 

For a simple plane wave solution we assume 
\beq
B_x(z,t)=Asin(kz-\omega t)
\eeq
which satisfies Eq. (8) with $\omega/k=v_A$.  For the velocity field Eq. (6) then yields
\beq
v_x(z,t)=-\frac{A}{\sqrt{4\pi \rho}} sin(kz-\omega t) .
\eeq
and from Eq. (5) the electric current is
\beq
J_y(z,t)=\frac{c}{4\pi} kA cos (kz-\omega t) 
\eeq
and the electric field is
\beq
E_y(z,t)=-A\frac{v_A}{c} sin(kz-\omega t) .
\eeq
The displacement of the fluid from an initial position $x_0$  is obtained from time integration of Eq. (11) and yields
\beq
x(z,t)=x_0 -\frac{A}{B_0k} cos(kz-\omega t) .
\eeq
where we have used that $B_0k=\omega \sqrt{4\pi \rho}$. 
On the other hand a magnetic field line at given time $t$,  $x_B(z,t)$, satisfies
\beq
\frac{\partial  x_B}{\partial z}=\frac{B_x(z,t)}{B_0}=\frac{A}{B_0}sin(kz-\omega t)
\eeq
from which we obtain
\beq
x_B(z,t)=x_B^0 -\frac{A}{B_0k}cos(kz-\omega t)  .
\eeq
Eqs. (14) and (16) show explicitly that magnetic field lines move together with the fluid, as of course they have to due to Alfven's theorem. Figure 3 shows the magnetic field lines,
fluid displacement, fluid velocity and electric current in such a wave. 

        \begin{figure} []
 \resizebox{8.5cm}{!}{\includegraphics[width=6cm]{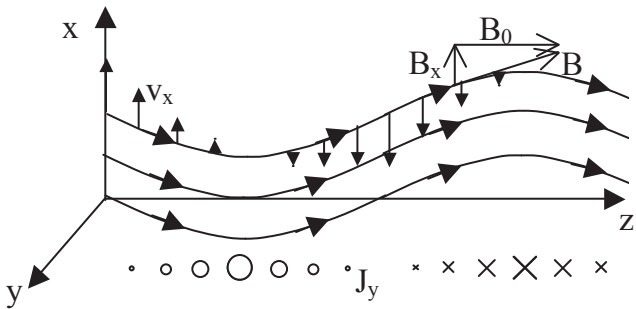}} 
 \caption {  Alfven wave discussed in the text. The wavy lines denote both position of fluid particles and magnetic field lines at 
 a given time $t$. As time evolves the lines move rigidly to the right with speed $v_A$, while fluid particles move up and down
 the $x$ direction as shown by the vertical arrows. The associated current $J_y$ pointing in (crosses) or out (circles) of the page is also
 indicated.   }
 \label{figure1}
 \end{figure}

\section{Type I superconductor in a magnetic field}
Type I superconductors in a magnetic field undergo a first order reversible phase transition for a critical
magnetic field $H_c(T)$ at temperature $T<T_c$, where $T_c$ is the critical temperature in the absence of magnetic field.
H.  London was the first to discuss the phase equilibrium between normal and superconducting phases in a seminal paper
\cite{londonh}. We consider a planar phase boundary for simplicity.

 Figure 4 shows coexistence of a superconducting (S) region for $x<x_0$ and a normal (N) region for $x>x_0$. 
 The plane defined by $x=x_0$ is the phase boundary. In the normal region, the magnetic field is
 \beq
 \vec{H}(x,y,z)=H_c \hat{z}  \; \;  \;  \; \;  \;   x\geq x_0
 \eeq
 and no current flows. In the superconducting region $x<x_0$  the magnetic field is
  \beq
 \vec{H}(x,y,z)=H_z(x)\hat{z}=H_ce^{(x-x_0)/\lambda_L} \hat{z}
 \eeq
 and a supercurrent
 \beq
 \vec{J}=J_y(x)\hat{y}= -\frac{c H_c}{4\pi \lambda_L}e^{(x-x_0)/\lambda_L} \hat{y}
 \eeq
 flows parallel to the phase boundary. Eqs. (18) and (19) follow from Ampere's law
 \beq
\vec{\nabla}\times\vec{H}=\frac{4\pi}{c}\vec{J}==>\frac{\partial H_z}{\partial x}=-\frac{4\pi}{c}J_y
 \eeq
 and the London equation
 \beq
 \vec{\nabla}\times\vec{J}=-\frac{c}{4\pi \lambda_L^2}\vec{H}==>\frac{\partial J_y}{\partial x}=-\frac{c}{4\pi \lambda_L^2} H_z .
 \eeq
The London penetration depth is given by \cite{tinkham}
 \beq
 \frac{1}{\lambda_L^2}=\frac{4\pi n_s e^2}{m^*c^2}
 \eeq
 where $n_s$ is the superfluid density and $m^*$ the effective mass of the charge carriers \cite{mstar}.
 The fact that normal and superconducting phases are in equilibrium at the phase boundary is determined by the fact that
 the free energy density difference between normal and superconducting phases, $F_n-F_s$,  is given by the kinetic energy density of the supercurrent \cite{londonh}:
 \beq
 \Delta F=F_n-F_s=\frac{H_c^2}{8\pi}=n_s\frac{1}{2}m^*v_s^2
 \eeq
 which follows from the fact that the superfluid velocity is given by
 \beq
 v_s=-\frac{e\lambda_L}{m^*c}H_c
 \eeq
 as determined by the London equation (21) with the supercurrent $J_y=n_sev_s$.

        \begin{figure} 
 \resizebox{8.5cm}{!}{\includegraphics[width=6cm]{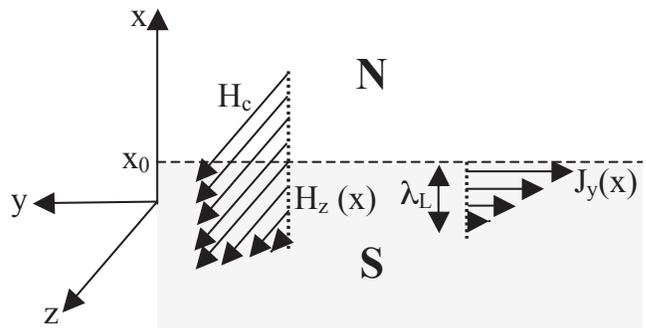}} 
 \caption { Normal (N, white) region for $x>x_0$ coexisting with superconducting (S, grey) region for $x<x_0$ in the presence of
 magnetic field $H_c$ pointing in the $z$ direction. Supercurrent $J_y$ flows in the $-y$ direction in a layer
 of thickness $\lambda_L$ in the superconducting region adjacent to the phase boundary, $x_0-\lambda_L<x<x_0$.   }
 \label{figure1}
 \end{figure} 

           \begin{figure} []
 \resizebox{8.5cm}{!}{\includegraphics[width=6cm]{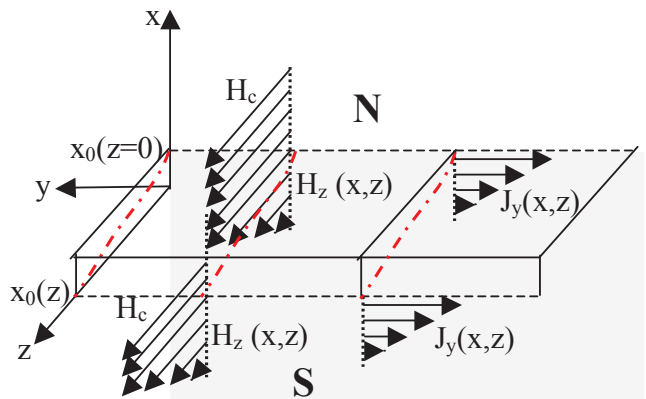}} 
 \caption { The phase boundary is now given by the function $x_0(z)$ indicated by the red dash-dotted line in the figure. 
 Now both $J_y$ and $H_z$ vary with $z$ also.  In addition, the magnetic field acquires a component in the
 $x$ direction, not shown in the figure. }
 \label{figure1}
 \end{figure} 
 
 Consider now a situation where we perturb the system so that the phase boundary is no longer perfectly planar. Rather than at a constant $x=x_0$, assume
 it is given by the surface $x=x_0(z)$. This is shown in Figure 5.  
 For example, that could be achieved by locally heating slightly so that the phase boundary deforms in the way shown in Fig. 5.

 In the superconducting region, the magnetic field and supercurrent are determined by Ampere's law and London's
 equation.
Let us assume that the supercurrent is given by the same form Eq. (19)
\beq
\vec{J}=J_y(x,z)\hat{y}= -\frac{c H_c}{4\pi \lambda_L}e^{(x-x_0(z))/\lambda_L} \hat{y} .
 \eeq
 The London equation now implies
 \bmath
 \beq
 \frac{\partial J_y}{\partial x}=-\frac{c}{4\pi \lambda_L^2} H_z
 \eeq
  \beq
 \frac{\partial J_y}{\partial z}=\frac{c}{4\pi \lambda_L^2} H_x
 \eeq
 \emath
 from which we deduce
 \bmath
 \beq
  H_z(x,z) =H_ce^{(x-x_0(z))/\lambda_L} 
 \eeq
  \beq
  H_x(x,z) =\frac{\partial x_0}{\partial z} H_ce^{(x-x_0(z))/\lambda_L} =\frac{\partial x_0}{\partial z} H_z(x,z)
 \eeq
 \emath
 
             \begin{figure} [t]
 \resizebox{8.5cm}{!}{\includegraphics[width=6cm]{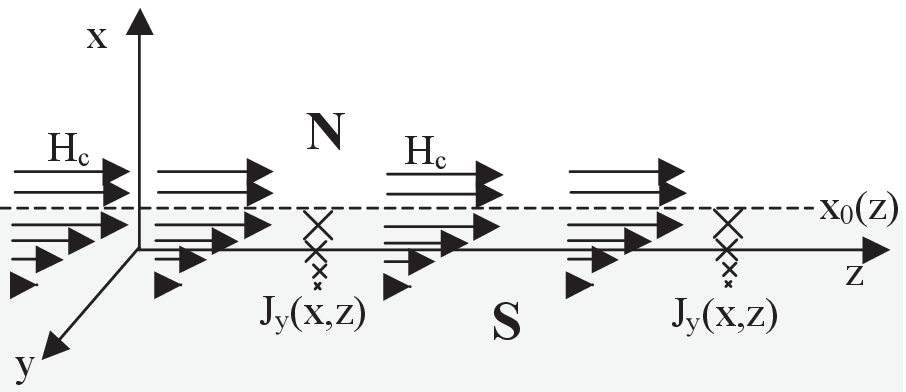}} 
 \caption { Figure 4 rotated by $90^o$ around the $x$ axis in counterclockwise direction. }
 \label{figure1}
 \end{figure}

 Now Ampere's law in this situation gives
 \beq
 \frac{\partial H_z}{\partial x} - \frac{\partial H_x}{\partial z} =-\frac{4\pi}{c}J_y
 \eeq
 while Eq. (27a) gives
 \beq
  \frac{\partial H_z}{\partial x}  =-\frac{4\pi}{c}J_y
  \eeq
  so the proposed solution is not valid unless we can neglect the second term in Eq. (28), which is
  \beq
  \frac{\partial H_x}{\partial z}=(\frac{\partial^2 x_0}{\partial z ^2} -(\frac{\partial x_0}{\partial z})^2 \frac{1}{\lambda_L})H_z .
  \eeq
  Therefore, our approximate solution requires that 
  \beq
\lambda_L \frac{\partial^2 x_0}{\partial z ^2} -(\frac{\partial x_0}{\partial z})^2  <<1 .
  \eeq
  In addition, the magnitude of the magnetic field at the phase boundary within this approximation is
  \beq
  H=H_c \sqrt{1+(\frac{\partial x_0}{\partial z})^2}\sim H_c(1+\frac{1}{2}(\frac{\partial x_0}{\partial z})^2)
  \eeq
  rather than $H_c$, so we also require that 
  \beq
  (\frac{\partial x_0}{\partial z})^2  <<1
  \eeq
  or in other words, both terms in Eq. (31) have to be separately small. This is not difficult to achieve, it   will be the case if the wavelength associated with the perturbation
  of the phase boundary is much larger than the amplitude of the phase boundary motion and than the London penetration depth.

             \begin{figure} [t]
 \resizebox{8.5cm}{!}{\includegraphics[width=6cm]{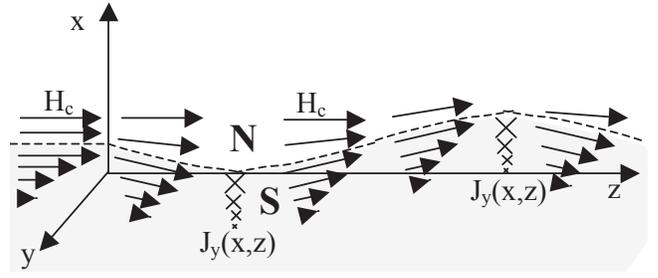}} 
 \caption { Figure 5 rotated by $90^o$ around the $x$ axis in counterclockwise direction. }
 \label{figure1}
 \end{figure}

Figures 6 and 7  show the magnetic fields and currents for the cases of figs. 4 and 5, rotated by $90^o$ around the
$x$ axis in counterclockwise direction. The dashed line gives the phase boundary. Note that Eq. (27b) implies
that the magnetic field lines in the superconducting region run parallel to the phase boundary.

  \section{interpretation of the results}
  
                \begin{figure} 
 \resizebox{8.5cm}{!}{\includegraphics[width=6cm]{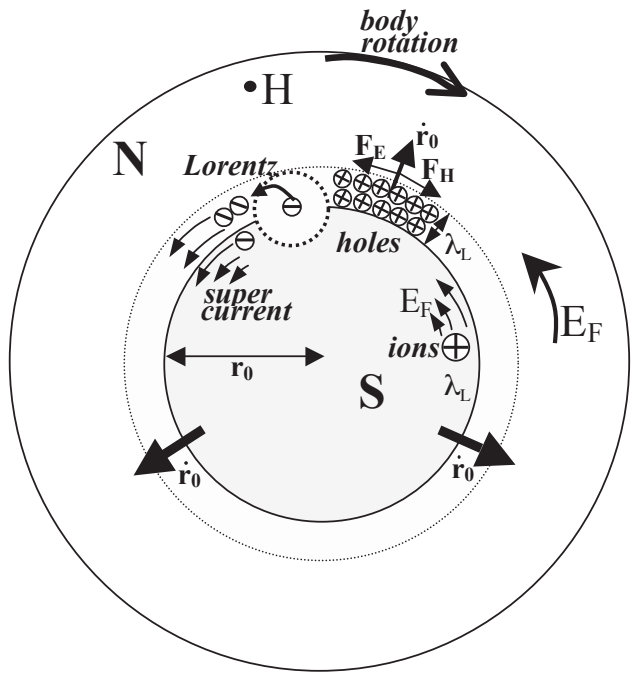}} 
 \caption { Meissner effect dynamics. As normal electrons become
superconducting, their orbits (dotted circle) expand, and the resulting Lorentz force propels the
supercurrent. An outflow of hole carriers moving in the same direction as the phase boundary
restores charge neutrality and transfers momentum to the body as a whole. See refs.  \cite{momentum, entropy}.}
 \label{figure1}
 \end{figure} 

Comparison of fig. 7 with the Alfven wave  depicted in fig. 3 strongly suggests that
the magnetic field configuration of Fig. 7 evolves from the magnetic field configuration of
fig. 6 by flow of a perfectly conducting   fluid in the x direction that carries the magnetic field lines of
fig. 6 with it, just as occurs in fig. 3. Note that the magnetic field gets modified up to a distance of order
$\lambda_L$ from the phase boundary, which implies that the flow has to reach a distance $\lambda_L$ within the superconductor.
Note also that the fluid considered in Sect. II was charge-neutral: Alfven waves don't propagate in non-neutral plasmas 
\cite{nonneutral}.

  Within conventional BCS theory, no charge flow in direction perpendicular to the  phase boundary is expected
  to occur when the phase boundary moves. We conclude that BCS theory cannot explain the dynamics of how the currents and fields
  change from fig. 6 to fig.  7.  
  
  More generally, the reader may argue that it is impossible to explain these results through flow of a charge-neutral fluid
  because there are no charge neutral fluids  in  solids, except in the very special case of compensated metals.

  Remarkably, the theory of hole superconductivity predicts precisely what is needed: that when the phase boundary
  moves, positive and negative charges move together with the phase boundary. This is shown schematically in fig. 8 for the case
  where the phase boundary moves into the normal region (Meissner effect). The same  motion in opposite direction occurs in the reverse process 
  when the phase boundary moves into the superconducting region.
  
  As discussed in the references and shown in fig. 8, within this theory electronic orbits expand from a microscopic radius to radius $2\lambda_L$ when they become
  superconducting, and the Lorentz force imparts on them the Meissner speed in counterclockwise direction. This is equivalent to electrons moving
  outward a distance $\lambda_L$. To compensate for this outward negative charge flow, there is a backflow of normal negative charge
  in the form of forward motion of positive charge, i.e. holes. As shown in the references, only if the normal electrons
have hole-like
  character, i.e. have negative effective band mass, it is possible to satisfy momentum conservation \cite{momentum}. It requires momentum transfer from
  the electrons to the body as a whole, which only carriers with hole-like character can do in a reversible way \cite{revers}, as required by
  thermodynamics \cite{entropy}.
  
  In a charge-neutral plasma or a liquid metal, the positive charges are ions and the negative charges are electrons. 
  Here, the positive charges are not ``real'' positive charges, they are a theoretical construct: $holes$. 
  But for the purposes of this paper we can forget that: as we learn in solid state physics textbooks \cite{am}, holes will
  behave just as real positive particles with positive mass under external fields, hence here they will behave   like the positive ions in
  a charge-neutral plasma. We don't need to consider here the issue of momentum transfer between electrons and the body that
 is important to understand the Meissner effect \cite{momentum}.
  
 For the purposes of this paper, what is important is that if our interpretation is correct, motion of the phase boundary is associated with
transport of mass, hence {\it there is inertia associated with motion of magnetic field lines}, just like in a plasma. There is also elasticity associated both with the 
energy of the magnetic field and the surface energy of the normal-superconductor boundary. This should allow for 
phase boundary wave propagation, as we discuss in the next section.

\section{phase boundary waves}

 As discussed in the previous sections, we expect that the dynamics of phase boundary motion in superconductors will  resemble
 the behavior of Alfven waves.  
 
 The speed of the Alfven wave Eq. (9) results from `elasticity' of the magnetic field lines. It can be understood from the fact that
 when the field line is not straight there is extra magnetic field energy density. In addition we need to consider the energy per unit area (surface tension) of the  interface which is given by   \cite{tinkham}
 \beq
 \sigma=\frac{H_c^2}{8\pi}\delta
 \eeq
 where 
 \beq \delta\sim \xi-\lambda_L, \eeq
  with $\xi$ the coherence length. When the phase boundary is not straight the 
 area of the interface increases and this gives rise to extra interface energy. 
The same happens for example at the interface between a liquid and a vapor phase, where the surface tension gives rise to capillary waves \cite{capillary}.
 As in that case, it will change the speed of propagation of the wave to
\beq
v_{pbw}=   \sqrt{\frac{H_c^2(1+ k \delta ) }{ 4\pi \rho} }
\eeq
where $k$ is the wavenumber. So it will be a small effect except for waves of wavelengths comparable to the coherence length or the London penetration depth.

Using $\rho=n_s m^*$ and Eq. (22) for the superfluid density, Eq. (36) yields
\beq
v_{pbw}=  \frac{eH_c}{m^*c}\lambda_L\sqrt{1+ k \delta} .
\eeq
or
\beq
v_{pbw}=0.176 H_c (G) \frac{m_e}{m^*}   \lambda_L(\AA) \sqrt{1+ k \delta}  \;   \;  cm/s
\eeq 
with $m_e$ the bare electron mass.

                  \begin{figure} 
 \resizebox{7.5cm}{!}{\includegraphics[width=6cm]{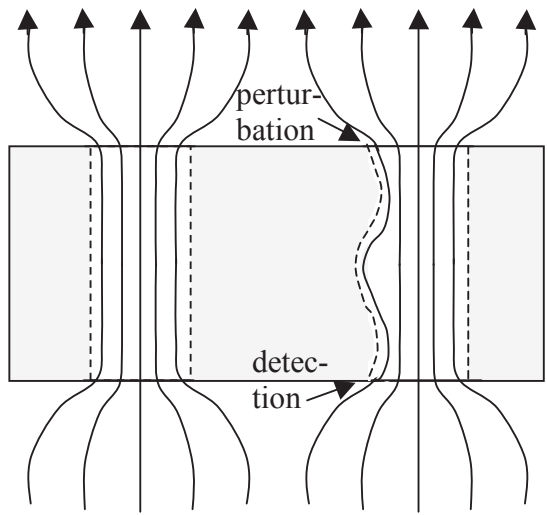}} 
 \caption { Type I superconductor in the intermediate state. The grey regions are superconducting, the white regions are normal. 
The upper surface is perturbed by a local magnetic field or temperature change, generating a  Alfven-like phase boundary wave.}
 \label{figure1}
 \end{figure} 
 
For example, assuming $m^*\sim m_e$ and $k\delta<<1$: for Pb, $H_c=803G$, $\xi=830\AA$, $\lambda_L=370\AA$ , $v_{pbw}=523m/s$; 
for Sn, $H_c=309G$, $\xi=2300\AA$, $\lambda_L=340\AA$, $v_{pbw}=185m/s$;
for Al, $H_c=105G$, $\xi=16,000\AA$, $\lambda_L=5000\AA$, $v_{pbw}=92m/s$.
These speeds are of the same order of magnitude as the speed of  sound   in air.

As we showed in ref. \cite{revers}, change in the magnetic field in the neighborhood of the phase boundary does not generate eddy currents and 
associated dissipation
$provided$ the normal charge carriers are holes. 
For this reason, for  small amplitude waves, meaning amplitude much smaller than $\lambda_L$, we expect that these waves should
propagate with essentially no damping, assuming the mean free path in the normal metal is of order of or larger than $\lambda_L$.
Therefore, these waves should propagate over macroscopic distances.

Associated with movement of the phase boundary there should be small changes in temperature due to
the latent heat of the transition: where the normal state grows the temperature will slightly drop, and vice versa.
Therefore, there will be an associated temperature wave analogous to second sound in  superfluid helium.

For example, in the geometry of fig. 9, with a superconducting slab of $cm$ dimensions in the intermediate state, a local perturbation at the upper surface 
next to a phase boundary wall should propagate down the slab and be detectable at the lower surface.
The perturbation could be a pulse or periodic, involving a change in the local magnetic field or the  temperature
that could be induced in a variety of ways with a local probe \cite{probes}.
With a pulse, the speed of propagation of the phase boundary wave should be measurable.

Instead, within the conventional theory of superconductivity, such waves should not propagate hence one would not detect
any effect at the lower surface for superconductors governed by the conventional theory when the upper surface
is perturbed.

There have been several experiments performed that detected Alfven waves in liquid metals \cite{exp1,exp2,exp3}.
For solids, Alfven waves have been detected in bismuth \cite{bismuth} and graphite \cite{graphite} that have equal density of electron and hole carriers hence constitute   neutral plasmas.
The experiments are difficult because of Joule dissipation and require either large magnetic fields or  large dimensions or both. 
For example, in ref. \cite{exp3} magnetic fields of up to $13T$ were used. For our case those constraints don't apply, 
detection should be feasible assuming  experimental capabilities to generate and detect small amplitude perturbations.

 \section{alfven's theorem in a conducting wire}
 Alfven's theorem states that in a perfectly conducting fluid, magnetic field lines are frozen in the fluid. When the fluid flows, it
 carries the magnetic field lines with it.
 It is a theorem, in the sense that it can be proven mathematically given  the laws of mechanics and electromagnetism \cite{davidson}.
 
We can also state Alfven's theorem in reverse: if it is observed that  magnetic field lines in a  conducting fluid move, 
 in the absence of external sources changing the magnetic field,  we can conclude that the motion of the
 magnetic field lines is caused by motion of the fluid. If the fluid is not perfectly conducting there will be dissipation  and magnetic field lines
 will only partially follow the fluid motion and will recede once the generated currents decay.

When metals in a magnetic field become superconducting, magnetic field lines move. The laws of mechanics and electromagnetism imply that
this motion is associated with motion of a conducting fluid. The conventional theory of superconductivity says there is no such motion.
Therefore it implicitly assumes that quantum mechanics somehow  cancels the effect of the laws of mechanics and electromagnetism. But it doesn't
explain how.

 The fact that magnetic field lines follow the motion of conducting fluids is vividly illustrated by a current-carrying superconducting wire.
 Figure 10 shows streamlines of electric current $\vec{J}$  for a wire that is superconducting in one portion and normal in another.
 There is electric charge flowing along the streamlines, with the velocity vector tangent to the streamline.
 The current distribution shown in Fig. 10 follows from solving London's equation and Ampere's law, as shown by London \cite{londonbook}.
 The current entering the superconductor spreads within a length $\lambda_L$ from the separating surfaces
 $z=\pm b$ toward the surface $r=a$ and then flows along this
 surface within a layer again of thickness $\lambda_L$ .

                  \begin{figure}
 \resizebox{8.5cm}{!}{\includegraphics[width=6cm]{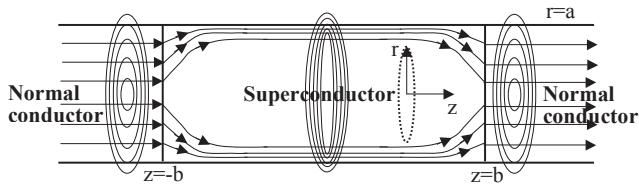}} 
 \caption { Streamlines of current flow in a cylindrical wire of radius $a$ that is superconducting in the region
 $-b<z<b$ and normal outside that region \cite{londonbook}, and associated magnetic field lines (circles).}
 \label{figure1}
 \end{figure} 
 
 It can be seen from  the picture, and can be shown mathematically, that the magnetic field lines in this system satisfy Alfven's theorem,
with  the electric current carried by a   conducting   fluid. The fluid carries the magnetic field lines
 with it, as if they were frozen  into  the fluid. In other words, the magnetic flux through any surface with its boundary attached to fluid particles   is preserved as the fluid 
 moves. When the fluid enters the superconducting region it flows to the top and bottom surfaces increasing its speed to satisfy the
 continuity equation, and carries the magnetic field lines along.
 
 How does the system reach the state of Fig. 10 if it is initially in the normal state carrying a current? 
 An intermediate step is shown in Fig. 11, where the current and magnetic field lines are partially expelled.
 The current streamlines move gradually towards the surface, which requires outward transfer of linear mechanical momentum, and they
 carry the magnetic field lines with them. 
According to the laws of mechanics and electromagnetism, this requires the flow of a conducting fluid from the interior
 to the surface as shown by the vertical arrows. Since no charge accumulates on the lateral surfaces,
 this outflowing conducting fluid has to be charge neutral. As in Fig. 8, we propose that it is composed of negative electrons and positive holes.
 
 Figs. 10 and 11 clearly illustrate that superconductors know about Alfven's theorem. We propose that in any situation where   a metal becomes superconducting and expels magnetic fields
 Alfven's theorem   holds. The expulsion of magnetic field from the interior   is associated with outward fluid motion.
 And more generally when the superconductor-normal phase boundary is displaced in any situation, since it results in motion of magnetic field lines,
 fluid motion in direction perpendicular to the phase boundary has to be involved.

 \section{discussion}
 
  It is  certainly true that physics at the microscopic level is different than at the macroscopic level. For example, Maxwell's equations dictate  that 
  electrons moving  in macroscopic  orbits radiate, and electrons in atomic orbits violate that dictum, because of quantum mechanics. 
  
  However, the converse generally does not apply. Macroscopic physical systems do not violate macroscopic laws of physics. 
  The only exception are superconductors according to the conventional theory of superconductivity: they  ignore Alfven's theorem when they expel magnetic fields.
However,  the conventional  theory  does not  explain how they achieve this.
  Instead, according to the theory of hole superconductivity, the expulsion of magnetic fields when 
  metals become superconducting is a manifestation of Alfven's theorem.

  Note that the electric field $E_y$ (Eq. 13) associated with an Alfven-like  wave along the phase boundary would also exist in the normal
  side of the phase boundary by continuity. One may expect that it will give rise to normal current flowing along the  phase boundary on the normal side
  that would give rise to dissipation and dampen the wave. However, we have shown in ref. \cite{revers} that because the normal state
  charge carriers are holes the tangential force due to $E_y$ is exactly cancelled by the magnetic Lorentz force acting in opposite direction when the phase boundary moves,
  so that no tangential normal current results and no eddy currents and dissipation occur. If the amplitude of the Alfven-like 
  wave is sufficiently large that magnetic field changes occur beyond a distance $\lambda_L$ of the phase boundary in the
  normal region,
  dissipation will set in. 
  
                         \begin{figure} [t!]
 \resizebox{8.5cm}{!}{\includegraphics[width=6cm]{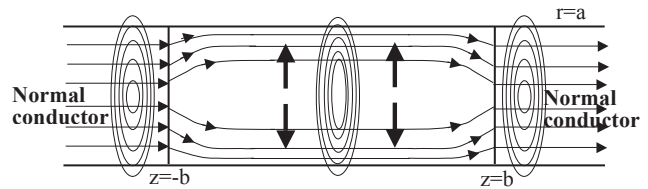}} 
 \caption { When the system enters the superconducting state
 current streamlines and magnetic field lines have to move outward  to reach the final state of fig. 10.
 It requires flow of a conducting fluid from the interior to the lateral surface as shown by the vertical arrows.}
 \label{figure1}
 \end{figure} 
 
    Experimental detection of the
  Alfven-like waves discussed in this paper 
  would be experimental proof that superconductors obey Alfven's theorem.
  Their existence would be evidence that there is inertia associated with
  motion of superconducting-normal phase boundaries. This requires flow of mass across the boundary when the
  phase boundary is displaced. The existence of these waves also requires that the flow of mass is not associated
  with a net flow of charge. In the normal state of solids, this is realized  for the cases of bismuth \cite{bismuth} and graphite \cite{graphite}, that have
  a compensated fluid of electrons and holes, where Alfven waves have been detected.   Here we have proposed that such waves can occur in any type I superconductor.
  It requires the normal state charge carriers to be holes, as predicted by the theory of hole superconductivity.
  For propagation over macroscopic distances the amplitude of the waves should be small compared to the London
  penetration depth.

  The fact that a rotating superconductor generates a magnetic field in direction parallel and not antiparallel to its
  angular momentum \cite{rotating} shows experimentally that the superfluid carriers in superconductors have negative charge.
  The same is shown by the gyromagnetic effect \cite{gyro} and the Bernoulli potential \cite{bern}. When the phase
  boundary moves, there is flow of both superfluid and normal fluid across the phase boundary. Thus, in order for this
  fluid flow to be charge neutral, since the superfluid flow is negative charge, 
  requires that the normal carriers are holes rather than electrons. This is perhaps the most compelling evidence in favor of the premise of the
  theory of hole superconductivity that only hole conductors can be superconductors.
  
  We were led to the prediction of these Alfven-like waves along normal-superconductor phase boundaries by 
  the prediction of the theory of hole superconductivity that there is charge flow and backflow \cite{ondyn}, and associated mass and momentum flow \cite{momentum}, across
  the phase boundary when it moves. 
  Instead, within the conventional theory of superconductivity there is no flow of charge nor of mass across the phase boundary
  when it moves, and the normal state charge carriers can be either electrons or holes. Nevertheless we can safely predict
  that when these waves are detected experimentally, many explanations will be ``found'' within the conventional theory,
  despite the fact that these waves were not predicted in the 60 years since the conventional theory was developed. 
Instead,   we argue that the detection of Alfven-like waves along normal-superconductor phase boundaries will  show   that the
  conventional  theory of superconductivity does not describe those superconductors.
  
  In this paper we have only considered type I superconductors. The possibility of Alfven-like waves in
  type II superconductors remains an open problem.

\end{document}